# GLOBAL MOBILITY AND HANDOVER MANAGEMENT FOR HETEROGENEOUS NETWORK IN VANET

Ravi Shankar Shukla[1], Neeraj Tyagi[2]

[1]Department of Computer Science and Engineering & IT, Invertis University, Bareilly, India
[2]Department of Computer Science & Engineering, MNNIT, Allahabad, India

## ABSTRACT

*Now a day's Vehicular Ad Hoc Network (VANET) is an emerging technology. Mobility management is one of the most challenging research issues for VANETs to support variety of intelligent transportation system (ITS) applications. VANETs are getting importance for inter-vehicle communication, because they allow the communication among vehicles without any infrastructure, configuration effort, and without the high costs of cellular networks. Besides local data exchange, vehicular applications may be used to accessing Internet services. The access is provided by Internet gateways located on the site of roadside. However, the Internet integration requires a respective mobility support of the vehicular ad hoc network. In this paper we will study about the network mobility approach in vehicular ad hoc network; the model will describe the movement of vehicles from one network to other network. The proposed handover scheme reduces the handover latency, packet loss signaling overhead.*

## 1. INTRODUCTION

The improvement of the network technologies has provided the use of them in several different fields. One of the most emergent applications of them is the development of the Vehicular Ad-hoc Networks (VANETs), one special kind of Mobile Ad-hoc Networks (MANETs) in which the communications are among the nearby vehicles. Now a day's vehicular Ad Hoc Network Communication is the wide area of research topic for Wireless technologies in education environment as well as automobile industry. Basically Vehicular Ad Hoc Network (VANET's) technically based upon the Intelligent Transportation Systems. Today mostly used mobility model are basically based on the simple random patterns model that cannot describe the vehicular mobility in the realistic way. Vehicular to vehicular (V2V) communication is the efficient due to various reasons like that short range, cheapest Communication and better bandwidth. VANET that is subclass of the MANET consist of the number of the vehicle travelling and communicating with each other without the fixed infrastructure this is the big benefit of the Ad Hoc Network that it is not required any fixed infrastructure. We can also characterized VANET on some other important basis i.e. high mobile node, potential large scale network and variable network density. VANET also provide facility for wide variety of applications for road safety, driving comfort, internet access and multimedia (Audio, Video). The ITS applications mostly on infrastructure based communications and used for the Internet Access by the Vehicles. In NEMO there ate multihop communication is not supported as they designed for the direct communication (single hop) with the Access Point. In this paper we present a Network Mobility Model for vehicular Ad Hoc Network and it is the scheme basically used wired cum Wireless Scenario. By the analysis of simulation, we show that our scheme provides better robust and seamless handover compared to other scheme.

## 2. RELATED WORK

Before the proposed handover scheme there are many model are introduced to reduce the packet loss and for reduce the latency. In order to support the real-time applications in Vehicular Ad-hoc Networks (VANETs), the proposal of a Leader-based scheme that exploits the topology of VANETs and a automatic configuration protocol like DHCP ensure the fast and stable address configuration. However, it assumes the use of a DHCP server and suffers from the control message overhead problem since it is a proactive protocol. Also, it needs the Duplicate Address Detection (DAD) when a vehicle changes its leader are describe in [4]. The optimized handover procedure of FMIPv6 by using Media Independent Handover (MIH) services in VANETs [6]. They introduced the information table to store the static and dynamic information of neighboring access networks and proposed to use a special cache maintained by the vehicle or by Access Router (AR) to reduce the anticipation time in FMIPv6, thus increasing the probability of the predictive mode of FMIPv6. The Global Mobility Management (GMM) was proposed for the inter-VANET handover of vehicles [10]. The proposed method supports the fast handover process using the L2 initiating and the route optimization for packet transmission. The Packet Forwarding Control (PFC) method was proposed in VANETs to select a Common Ahead Point (CAP) to forward packets [3]. The MMIP6, a communication protocol that integrates multi-hop IPv6-based vehicular ad hoc networks into the Internet [8]. MMIP6 is highly optimized for scalability and efficiency based on the principles of Mobile IPv4 and does not provide the interoperability with IPv6, FMIPv6 and HMIPv6 mobility schemes. A NEMO protocol proposed for VANETs [9]. The major problem happen in MIPv6 is the handover latency, the time interval in which Vehicular Node (VN) cannot send or receive packets during handover.

## 3. Proposed Network Architecture

In the proposed architecture, heterogeneous networks are considered where a mobile network is provided with mobility through a number of ISPs (Internet Service Provider) as shown in the figure 1. The MRs (Mobile Routers) are associated with ARs (Access Routers) of different access network e.g. WLAN network and cellular network. The ISPs have been assigned different home addresses and home agents to the vehicle equipped with MRs and each MR configures a global IPv6 address through router advertisement from the AR. During handover process, flow redirection may create a problem when traffic is redirected between MRs in different administrative domains. In such cases ingress filtering prevents MR from communicating with home network of other MRs because MRs belong to different HAs. Thus, the MR is now forced to use a different home address which can cause the termination of ongoing communication sessions.

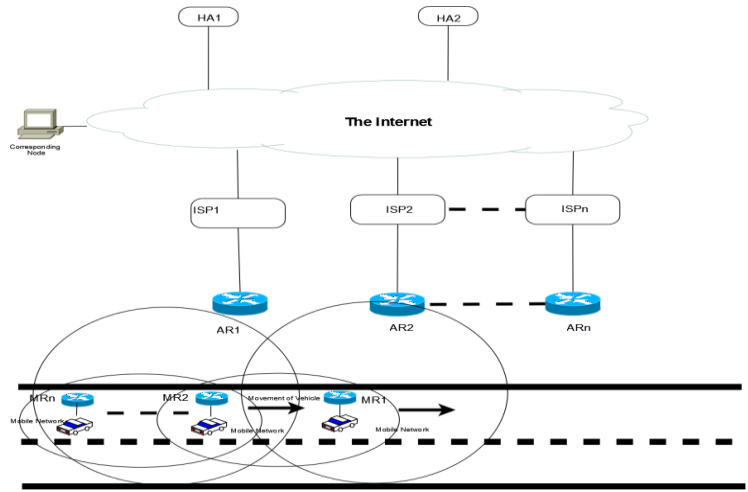

Figure 1: Proposed Network Architecture

The assumption made in proposed architecture assumes the following parameters:

(1) Range of AR ($AR_i$) is fixed (i.e. 100 m radius) for better communication.

(2) Distance between two AR's ($d_{AR}$) is assumed to be 150 m so the maximum overlapping region between two AR's is 50 m on the road as shown in figure 1.

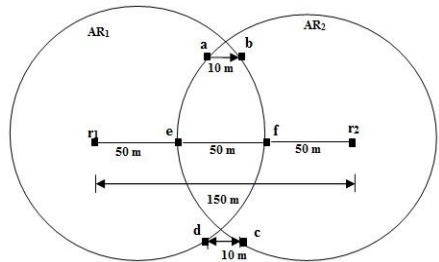

Figure 2:

(3) It is also assumed that the road topology & location of AR's on roadside is fixed, such that the minimum and maximum length of overlapping region is 10 m and 50 m respectively. It is justified by the argument that the width of roads is in the range of 40-50m in highway as shown in figure by point a, b, c, d.

(4) Another assumption which is also considered is that the traffic is on the road is sufficient enough that communication between two vehicles with MR's is always possible with fixed range of MR's, it also implies that vehicles are running nearly in the same speed.

The proposed architecture can support any number of vehicles and every vehicle must equipped with Mobile Router (MR). For vehicular scenario, we can assume any number of vehicles, but for the alliterations purpose, we have considered two vehicles that are deployed on the road and every vehicle maintains an ad-hoc network communication. The figure 2 represented a details view of handover region as shown in figure 1. In figure 3, MR1 and MR2 are mounted on first and second vehicles respectively. When vehicles move from the subnet of AR1 to AR2, vehicle equipped with MR1 leaves the AR1 and vehicle equipped with MR2 is still in the range of AR1, so that vehicle connectivity of MR1 to the internet is maintained through by MR2 until MR1 completes handover process. During handover period the traffic addressed to MR1 is transported via MR2. The vehicle gets internet connectivity through different ARs (AR1, AR2….,$AR_n$) via MR2 and MR1 respectively for some time until MR2 leaves AR1. During MR2's handover period the traffic addressed to MR2 is transported by MR1.

Let $A_{(x1, y1)}$ or $B_{(x2,y2)}$ be any point on the boundary of overlapping region of AR1 or AR2 such that MR1 enters or exit from point A or B. (See figure 2), for simplicity, we write $A_{(x1, y1)}$ as A and $B_{(x2,y2)}$ as B.

According to the assumption number (3) the distance travelled by any vehicle in the overlapping region lies between 10 m $\leq$ d (A, B) $\leq$ 50 m.

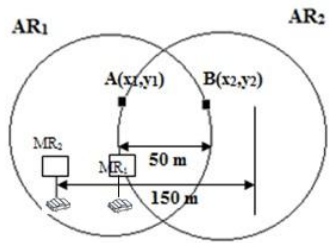

Figure 3:

When MR1 comes into the overlapping region then it gets weaker signal and hence disconnection problem with the internet may arise. But it can be overcome by the fact that enough traffic exists on the road for communication among the vehicles. This implies that MR1 remains connected to AR1 since MR2 is still in the range of AR1 and there is ad hoc communication between MR1 and MR2. At the same time, the $MR_1$ sends message to $AR_2$ for registration. As long as the registration process is not completed, MR1 gets the message from $MR_2$. The time duration for $MR_1$ to complete handover process is given by:

$$t_{hp} = \frac{d(A,B)}{V_h} = \frac{distance\ traveled\ handover\ region}{Vehicle}$$

Or    $t_{hp}$ =distance travelled in handover region / vehicle speed where d(A,B) is the distance between overlapping region at point  A & B which lies at boundary region & $V_h$  is the speed of vehicle (MR1)

Formally,

$t_{hp}$ is minimum when d(A,B)=10 m &
$t_{hp}$ is maximum when d(A,B)=50 m

assuming $V_h$ to be constant in both cases

So

$$\frac{10}{V_h} \le t_{hp} \le \frac{50}{V_h}$$

## 4. Proposed Handover Scheme and Tunneling Establishment:

To achieve seamless handover of vehicular network, the proposed scheme is to redirect the traffic flow of the MRs (which is mounted on the vehicles) to one another via different HAs during handover period. When the mobile network moves along with the vehicle, the Mobile Router1 (which is installed on first vehicle V1) detects AR2 through Router advertisement message from AR2. As it moves from the subnet Access Router1, it loses the connection from the global Internet. The traffic of MR1 should be transported through HA1-MR2 (which is installed on vehicle V2) using the bi-directional tunnel until it completes handover to AR2. When MR1 moves out of the subnet of AR1and undergoes handover process, it re-establishes the tunnel with HA1 and executes the following functions:

### (i) Tunneling with AR1:

MR1 informs HA1 by Binding Update (BU) message via MR2 to tunnel its packet to MR2, which is still in the subnet of AR1, as MR1and MR2 are both in the ad-hoc network region. Figure 4 shows the tunnel established of MR2 during its handover period. The home address, destination and alternate care-of-address option field of the binding update message contains MR1 previous care-of-address (IP2) and MR2's previous care-of-address (IP1) respectively. On receiving BU message, HA1 creates a mapping between IP1 and IP2 which is used to tunnel the packets addressed to IP2 through AR1. All the packets routed to MR1 are intercepted by HA1 and the packets are encapsulated and forwarded to IP1. The encapsulated packets routed to MR1 and decapsulated to take IP1 off. The destined address is IP2 which is send by the Correspondent Node (CN).

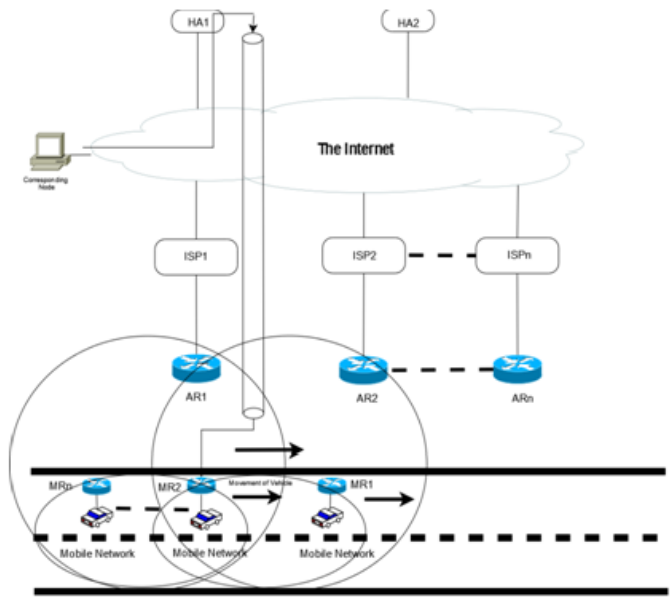

Figure 4: Tunnel establishment of MR1 during its handover

## (ii) New CoA configuration and HA2 registration:

MR1 sends a route solicitation message for configuring a new CoA (IP3) to AR2. MR1 receives an route advertisement message broadcasted by AR2 and configures IP3. MR1 verifies the uniqueness of IP3 by sending **NA** message to AR2 and simultaneously receiving the NAck message. After IP3 configuration, MR1 deletes the previous Home Agent (HA1) record and perform the registration of a new home agent (HA2) by sending a new binding update (HA2-BU) message to HA2. When HA2 replies with a BAck message, it tunnels packet to IP3 through AR2.

After completing the handover process, MR1 reaches in the service area of different service provider (ISP2) using different access technology than ISP1. ISP2 assigns different home address and home agent to MR1. The mobile network is now connected with different home networks through MR2 and MR1 respectively, i.e. MR2 and MR1 are advertising different prefixes on their ingress interfaces. The prefix advertised by MR2 is registered to HA1 and prefix advertised by MR1 is registered to HA2. Once MR1 has completed the handover process and started receiving packets through AR2, now Vehicle equipped with MR2 experience that it has to perform handover to AR2 in the near time. MR2 can start establishing a new tunnel using alternative route, this action causes MR2 to perform handover with a little disruption as possible. MR2 then inform HA1 to tunnel its packets through HA2-MR1 bi-directional tunnel to MR1 which is located in the subnet of AR2. In this case, if HA2 performs ingress filtering packets with a source address prefix of MR2 may be discarded.

## (iii) Tunneling with AR2:

MR2 can prevent ingress filtering from dropping the packets when the two tunnels end at different HAs by re-establishing the tunnel with HA1 through CoA configured from the MNP advertised by MR1. Figure 5 shows the bi-directional tunnel established by the new binding and for doing so MR2 executes the following functions:

1. MR2 gets its own ingress interface along with a new CoA with the prefix announced by MR1. Then, it sends BU message to HA1 using new CoA. On receiving BU message, HA1 can encapsulate incoming packets through the bi-directional tunnel via MR1.

2. All the packets routed to MR2 are intercepted by HA1. The new binding causes HA1 to tunnel the packets with new CoA of binding cache in HA1 to the MR2. Thus the packets are encapsulated and forwarded to the new CoA of MR2 via HA2-MR1 bi-directional tunnel.

After successful BU, all the traffic between CN and MR will experience the two levels of encapsulation. The first level of tunneling is done between MR2 and HA1 and the second level of tunneling is done between MR1 and HA2. This mechanism allows MR1 to send and receive packets using MR2's bi-directional tunnel and there is no need for MR2 to change HoA and ongoing sessions do not cause disruption during handover.

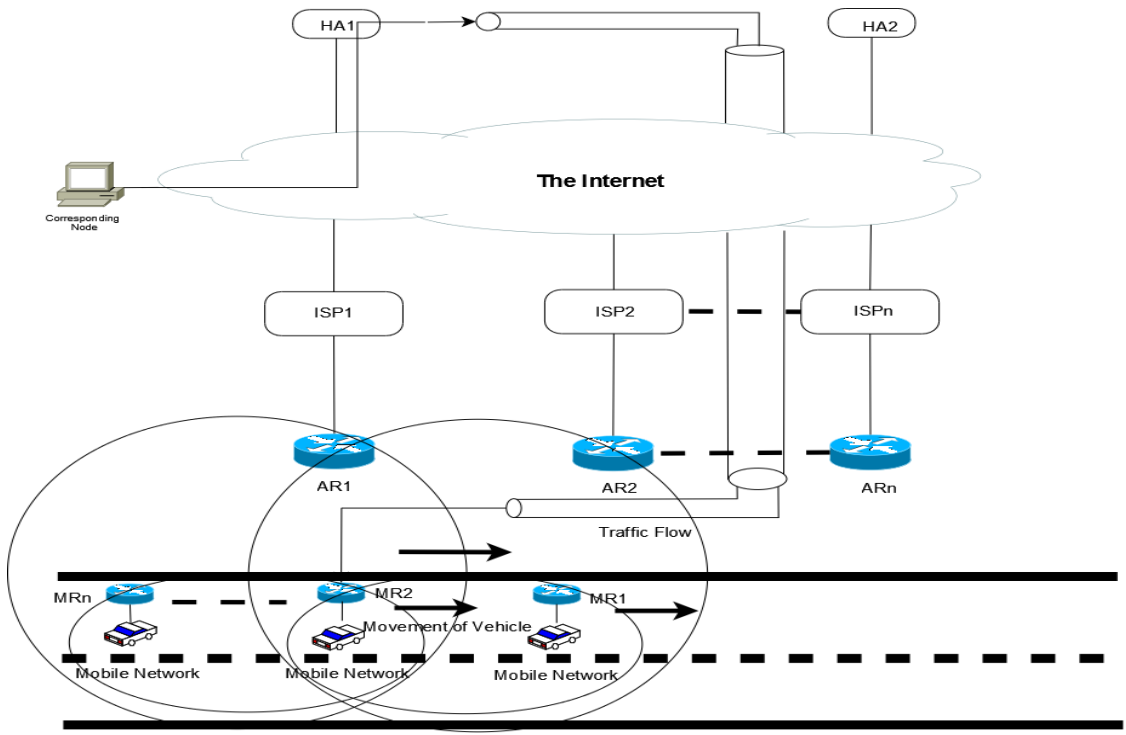

Figure 5: Tunnel re-establishment of MR2 during its handover

**(iv)** Depending upon the availability and strength of the radio signal from AR2, MR1 decides whether to stay in AR1 or move to the AR2. During handover to AR2, MR2 executes the same sequence of functions as performed by MR1.

The same process will perform in other mobile vehicles which are on the road and try to get global internet connectivity, through different ARs (AR1, AR2….,ARn).

## 5. SIMULATION PARAMETER AND RESULTS ANALYSIS

| Parameters | Values |
|---|---|
| Simulation Area | 10,000*40 m |
| Simulation Time | 10 ms |
| ISP Coverage | 0-10,000 m |
| Gap between ISP Range | 150 m |
| Required Interval | 1s |
| Number of Vehicles | 30 |
| Vehicles Speed | 50 Kmph to 120 Kmph |
| Ad hoc Data Transfer Rate | 4 Mbps |
| Packet Size | 400 Bytes |
| Ad hoc Coverage | 900 m |

Table 1. Simulation Parameter

To evaluate the performance of our proposed scheme and FMIPv6, simulation was performed at various parameters as shown in Table 1. We use several abstract function in simulation topologysuch as HA, CN, FA, AR and Vehicular node (VN).

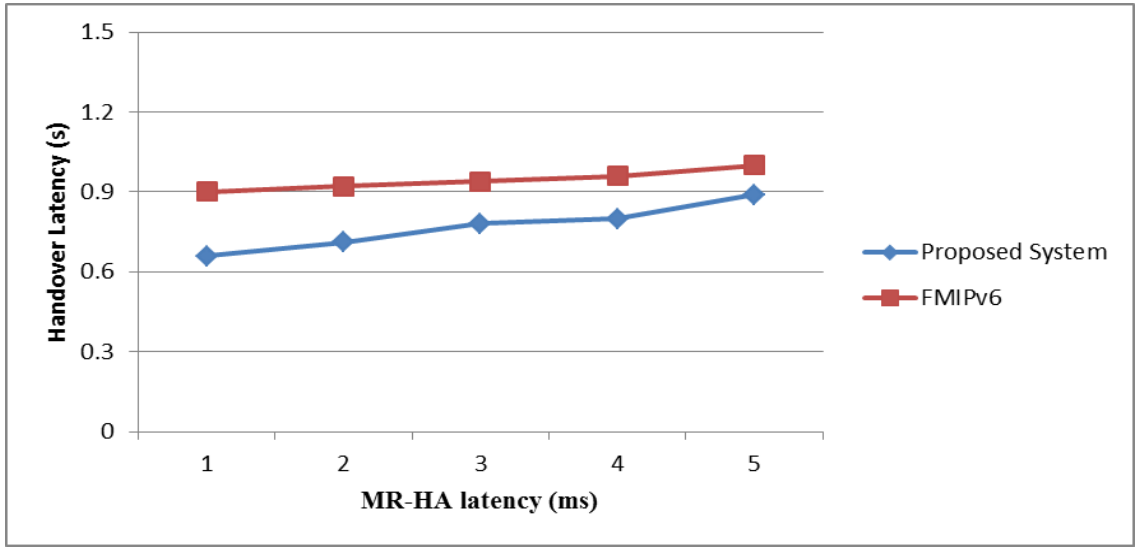

Figure 6:Handover Latency

## 5.1 Handover latency

Figure 6 shows the handover latency experienced during the simulation of proposed scheme and is compared with FMIPv6. Handover latency of a mobile network is defined as the complete handover time from one access router (AR1) to another access router (AR2) and it is proportional to the distance between MR and HA for all the protocols. However, the latency of the proposed scheme is less than FMIPv6 because FMIPv6 requires a number of signaling messages in establishing tunnel between AR1 and AR2 before performing HA-BU. Another reason of less handover latency is that the second vehicle MR (MR2) can configure its new CoA in advance by getting Rt Adv information from MR1 for actual handover.

## 5.2 Service disruption time

In FMIPv6 service disruption time during handover can be defined as the time between the receptions of last packet from AR1 until the first packet is received from AR2 Via the tunnel established between AR1 and AR2. In the proposed scheme, service disruption time during handover can be defined as the time between the reception of the last packet through the MR which is about to undergo handover process until the first packet is received through another MR of the mobile network. Figure 7 compares the service disruption time between the proposed scheme and FMIPv6 with respect to MR-HA latency. The disruption time of FMIPv6 is about 0.9 second (because of the signaling and time it takes in establishing tunnel between AR1 and AR2), while the service disruption time of the proposed scheme is close to 0.2 second (because only one signaling message is required to reestablish the bi-directional tunnel). This enhances that the proposed scheme can support seamless network mobility.

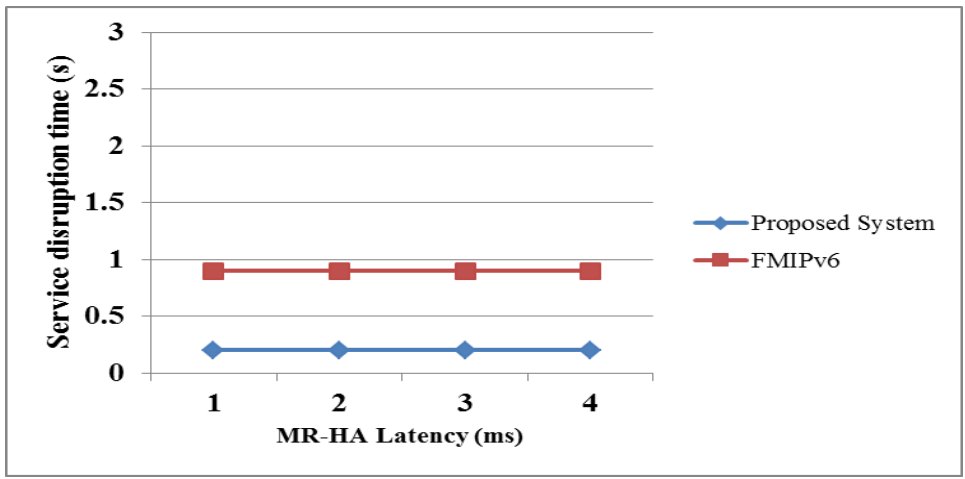

Figure 7: Service disruption time

## 5.3 Packet loss

Figure 8 shows the packet loss during the handover from AR1 to AR2. It is evident from the graph that lowest number of packet loss is experienced in the case of proposed scheme because of the cooperative packet reception of MRs. The packets addressed to the MR which is undergoing handover process are received via another MR by reestablishing the bi-directional tunnel and the packet loss is independent of the distance between MR and it's HA. FMIPv6 also establishes a tunnel between AR1 and AR2 and the packet loss is independent of MR-HA latency. Due to the number of signaling messages involved in establishing the tunnel, the number of lost packets is higher than that of the proposed scheme.

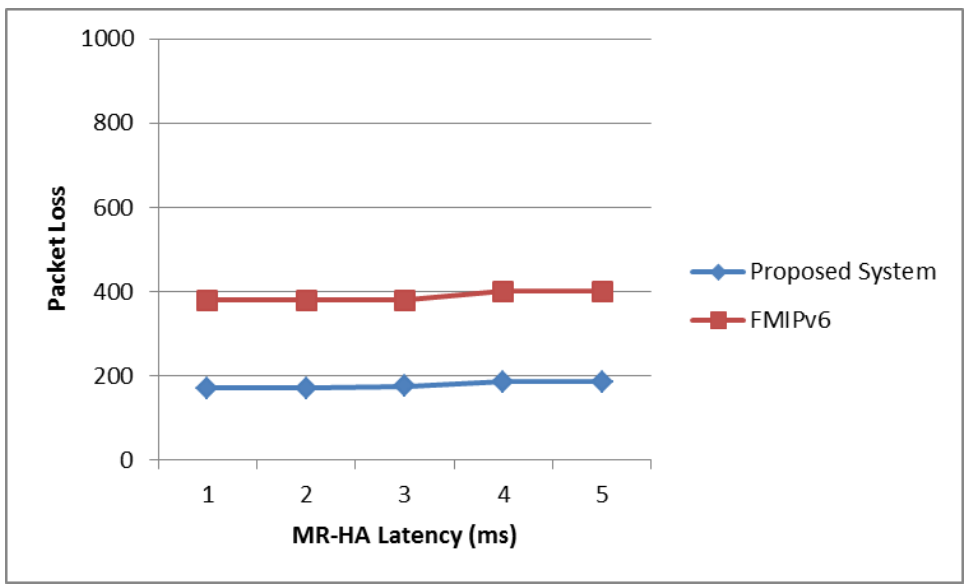

Figure 8: Packet loss

## 5.4 Signaling overhead

Figure 9 compares the signaling overhead ratio between FMIPv6 and the proposed scheme in multiple MRs based mobile network. Signaling overhead involves the exchange of signaling messages to manage handover process effectively, such as the messages required for re-establishing the bi-directional tunnel to achieve flow redirection of MRs via one another. The proposed scheme has some overhead because the cast involves in maintaining signaling messages of MRs. However, it has less overhead in comparison to FMIPv6, because FMIPv6 involves a lot of signaling messages to establish a tunnel between AR1 and AR2. The proposed scheme increases the signaling overhead in the network and signaling overhead of these protocols increases with increasing MR-HA latency.

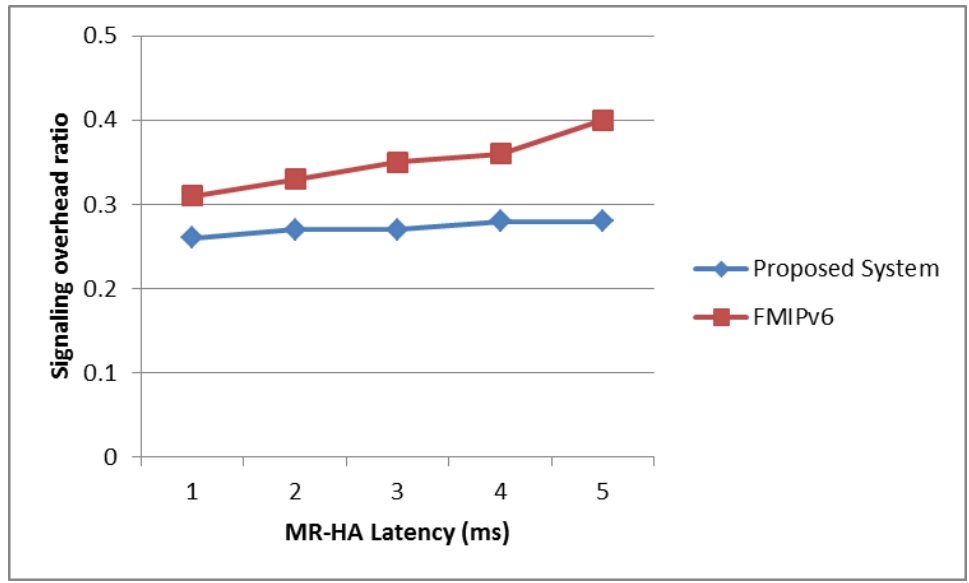

Figure 9: Signaling Overhead ratio

## 6. CONCLUSION

The proposed architecture support seamless mobility of mobile network across heterogeneous network. A vehicular scenario is considered where a vehicle is provided with mobility through different Internet Server Protocol's (ISPs) and purpose to use multiple Mobile Router s (MR) based handover schema in vehicle. The multiple Mobile Router (MR) based handover schema where MRs cooperatively receive packet destined for each other can provide no service disruption and significantly reduce packet loss during handover. It also makes the packet loss independent to handover latency. Moreover Multiple Mobile Router's (MR's) architecture is extended to include Multiple Home Agent's, where each Home Agent belongs to different administrative domains. This endows the vehicle to be able to do smooth handover over heterogeneous network where mobility is provided through different Internet Server Protocols (ISP's). An extensive simulation study is carried out to show the comparative performance evolutions of the proposed handover architecture in terms of throughput, handover latency service disruption time, packet loss and signaling overhead. The simulation results shows that the proposed architecture provide a mobile network with seamless mobility across heterogeneous networks. The overlapped reception of packet from different Access Router's (AR's) significantly minimizes packet losses during handover even without reducing handover latency.

**Authors**

Ravi S Shukla obtained a M.Tech. Degree in Software Engineering in 2002 from MNNIT, Allahabad, Deemed University. Currently, he is working as an Associate Professor in the Dept. of CSE & IT and research scholar of MNNIT, Allahabad in the dept. of CSE. His research interest includes Computer Networking, TCP/IP, Mobile Computing and Adhoc networking. 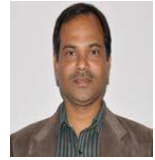

Dr Neeraj Tyagi obtained a Ph.D. degree in 2008 in Computer Science & Engineering from MNNIT, Allahabad, Deemed University. Currently, he is working as an Associate Professor in the Dept. of CSE. His research interest includes Computer Networking, TCP/IP, Mobile Computing and Adhoc networking. 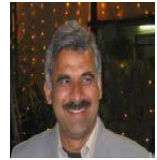